# On the determination of the thermal shock parameter of MAX phases: A combined experimental-computational study

Matej Fekete[1*], Clio Azina[1], Pavel Ondračka[1,2], Lukas Löfler[1], Dimitri Bogdanovski[1], Daniel Primetzhofer[3], Marcus Hans[1], Jochen M. Schneider[1]

[1]Materials Chemistry, RWTH Aachen University, Kopernikusstr. 10, D-52074 Aachen, Germany

[2]Department of Physical Electronics, Faculty of Science, Masaryk University, Kotlářská 2, 611 37 Brno, Czech Republic

[3]Department of Physics and Astronomy, Uppsala University, Lägerhyddsvägen 1, S-75120 Uppsala, Sweden

[*]Corresponding author’s e-mail address: fekete@mch.rwth-aachen.de, fekete@physics.muni.cz

**Abstract**

Thermal shock resistance is one of the performance-defining properties for applications where extreme temperature gradients are required. The thermal shock resistance of a material can be described by means of the thermal shock parameter $R_T$. Here, the thermo-mechanical properties required for the calculation of $R_T$ are quantum-mechanically predicted, experimentally determined, and compared for $Ti_3AlC_2$ and $Cr_2AlC$ MAX phases. The coatings are synthesized utilizing direct current magnetron sputtering without additional heating, followed by vacuum annealing. It is shown that the $R_T$ of both $Ti_3AlC_2$ and $Cr_2AlC$ obtained via simulations are in good agreement with the experimentally obtained ones. Comparing the MAX phase coatings, both experiments and simulations indicate superior thermal shock behavior of $Ti_3AlC_2$ compared to $Cr_2AlC$, attributed primarily to the larger linear coefficient of thermal expansion of $Cr_2AlC$. The results presented herein underline the potential of *ab initio* calculations for predicting the thermal shock behavior of ionically-covalently bonded materials.

## Introduction

It has long been known that materials used in demanding applications where rapid and large temperature changes occur, such as in jet or rocket propulsion components, internal combustion engines, metallurgical processes, or nuclear waste storage, may undergo catastrophic failure [1,2]. This behavior is caused by abrupt temperature changes resulting in very steep thermal and stress gradients, which in turn lead to rapid material expansion and resulting deformation, crack formation, and, finally, catastrophic failure – a phenomenon known as thermal shock [1–3]. Enhancing the thermal shock resistance of a given phase mitigates this vulnerability and is thus essential for the design of high-performance materials.

Expressed in terms of dependence on key material properties, multiple definitions of the thermal shock resistance parameter ($R_T$) exist, with a common one being [2,4]:

$$R_T = \frac{\sigma_f \kappa (1-\nu)}{\alpha E}, \tag{1}$$

where $\sigma_f$ is the flexural strength, $\kappa$ is the thermal conductivity, $\nu$ is the Poisson's ratio, $\alpha$ is the linear coefficient of thermal expansion (LCTE), and $E$ is the elastic modulus. From this definition, it is evident that for maximal thermal shock resistance, the flexural strength and thermal conductivity should be high (to resist deformation and reduce thermal gradients), while the LCTE and elastic modulus should be low (reducing the stress gradient). However, a major problem is that frequently, improving one of these properties leads to concurrent deterioration of another, e.g., the increase of the LCTE when the elastic modulus is minimized [5]. In addition, these properties in turn depend on multiple extrinsic factors: for instance, the flexural strength, thermal conductivity, and LCTE depend on grain size [6–8], while residual stress may affect the elastic modulus [9], potentially reinforcing -or counteracting- the induced thermal stress. In summary, the efficient design of materials with a high $R_T$ is a multivariate optimization problem, with all challenges that this entails; in practice, often the optimization of one particular property, such as, e.g., the thermal conductivity, is pursued.

A further complicating aspect arises in the case of thermal shock-resistant coatings, where the role of the substrate needs to be considered as well, as thermal shock resistance can degrade in case of a substrate-coating thermal expansion mismatch [10]. This issue is pertinent as such coatings are increasingly used to protect components in demanding applications (e.g. in gas turbines or combustion engines) [11,12]. Even more essential in such applications is a high thermal conductivity to dissipate strong heat buildup and thus increase the thermal shock resistance, as evident from equation (1). A highly promising class of materials for use in such coatings are nanolaminated multielement compounds known as MAX phases, in which M is a transition metal, A is an element from group IIIA or IVA, and X is usually C and/or N; the common structure is that of transition metal carbide/nitride trigonal prisms separated by layers of the A element [13]. These materials are distinguished by good mechanical properties such as high elastic moduli, good machinability, and high damage tolerance [13–15] with simultaneously high thermal stability and good thermal conductivity [14,16], making them resistant to thermal shock and mechanically well-suited for demanding applications.

The good thermal shock resistance of MAX phases has been highlighted by several authors [6,17–20]. El-Raghy *et al.* reported that the thermal shock resistance of bulk $Ti_3SiC_2$ was dependent on the microstructure, as the fine-grained $Ti_3SiC_2$ specimens were susceptible to thermal shock while the coarse-grained ones were not [6]. Further studies were then carried out to describe the abnormal thermal shock behavior of MAX phases. Indeed, the residual strength of MAX phases post-quenching in water tends to decrease with increasing quenching temperature up to a critical temperature, after which the residual strength increases. Bao *et al.* reported the retained strength of $Ti_3AlC_2$ after quenching in different media at different temperatures [18]. They noticed that quenching temperatures between 300 and 500 °C led to strength loss while quenching temperatures between 1000 and 1300 °C, in contrast, led to strength increase [16]. Aside from $Ti_3AlC_2$ [18,19,21], similar observations were also reported for $Ti_2AlC$ [17,20], $Ti_3SiC_2$ [6,19], and $Cr_2AlC$ [22,23]. The latter is another particularly prominent MAX phase, first synthesized in thin film form by Schneider *et al.* [24] with a magnetron-sputtering-based deposition approach. In a later study, deposition of a stable, crystalline $Cr_2AlC$ phase on a steel substrate at a temperature of 450 °C was shown to be possible, governed by surface diffusion of the deposited atoms [25], enabling a pathway towards large-scale deposition on technologically relevant substrates. In line with the general characteristics of MAX phases, $Cr_2AlC$ also was shown to exhibit a high elastic modulus as determined from measurements and *ab initio* calculations of the elastic tensors in multiple studies, reported in the range between 138 [26,27] and 228 GPa [24,28,29]. Using such a correlative theoretical and experimental approach, very good agreement between measured and predicted elastic moduli was obtained for sufficiently phase-pure and dense films [29]. Generally speaking, *ab initio* calculations utilizing density functional theory (DFT) have been used to elucidate further properties of $Cr_2AlC$ and other MAX phases as well, e.g., correlating the ductile behavior of $Ti_2AlC$, $Cr_2AlC$, and $V_2AlC$ with the electronic structure of the respective systems, predicting that the weaker, nondirectional metal-metal bonds govern the ductility [30].

Thus, both measurements and predictive calculations of material properties are essential to understand the complex interplay of various factors, which is particularly important in the case of thermal shock resistance due to the outlined distinct, and occasionally counteracting, properties governing the overall response. To that end, in the present work, we examine the thermal conductivity, flexural strength, linear coefficient of thermal expansion, and elastic modulus of two MAX phases, $Ti_3AlC_2$ and $Cr_2AlC$, deposited via magnetron sputtering, using a combined experimental (nanoindentation and X-ray diffraction during *in-situ* annealing) and theoretical (calculation of elastic tensors and resulting moduli as well as predictions of the thermal conductivity) approach. To our knowledge, this is the first contribution examining this array of properties specifically with the aim of gauging their impact on the thermal shock parameter and, in the long run, enabling controlled tuning of this important property.

**Experimental details**

A high-vacuum combinatorial physical vapor deposition system with a base pressure below $5\times10^{-5}$ Pa was utilized to synthesize both coatings. The deposition of the Ti–Al–C film was performed using elemental 2-inch Ti, Al, and C targets (99.9 % purity), utilizing conventional direct current (DC) magnetron sputtering at a constant power of 130, 45, and 190 W, respectively. In the case of Cr–Al–C, DC magnetron sputtering of a composite Cr–Al–C (2:1:1 composition) 2-inch target (Plansee Composite Materials, Germany) at a constant power of 200 W was employed. In both cases, Ar was used as working gas, with a deposition pressure of 0.4 Pa for Ti–Al–C and 0.45 Pa for Cr–Al–C. Schematic diagrams of the respective deposition setups are shown in Figure 1 left (Ti–Al–C) and right (Cr–Al–C). All films were deposited on MgO (100) substrates placed 10 cm from the target and rotated at a constant speed of 10 rpm. The substrates were kept at floating potential and were not intentionally heated.

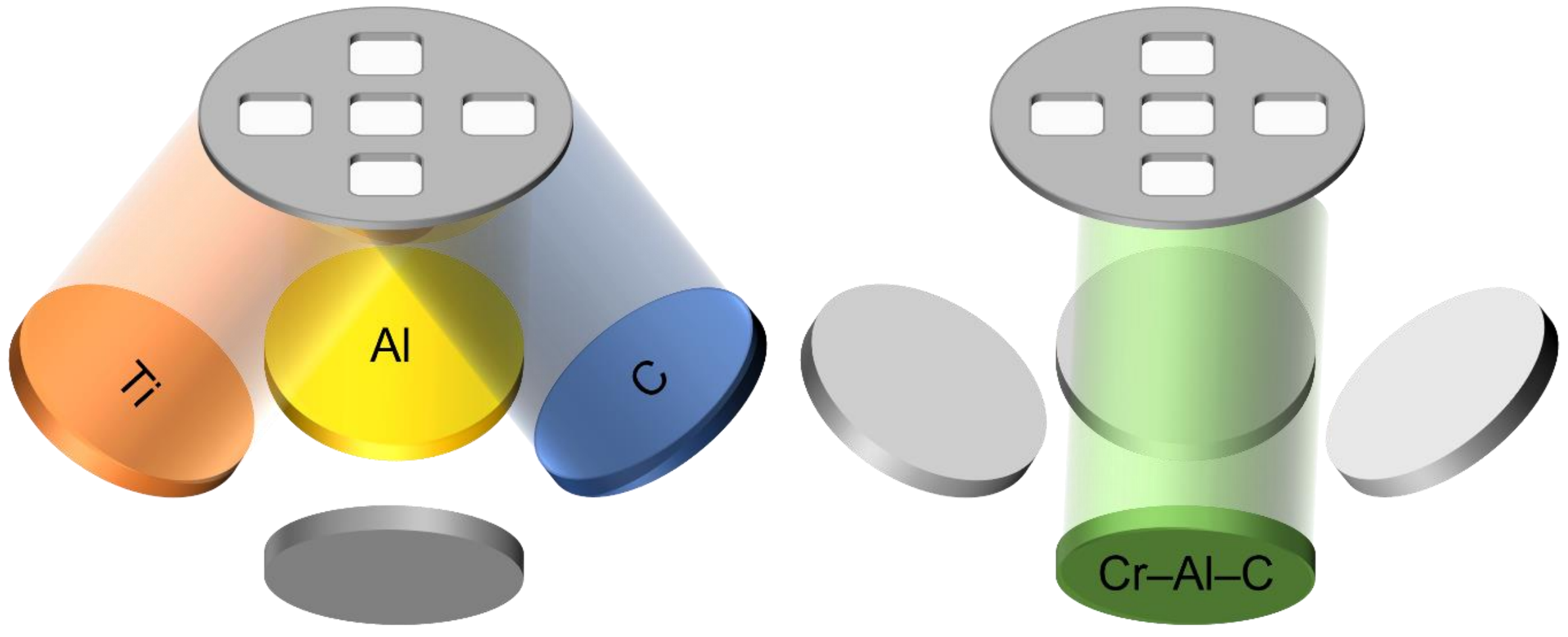

*Figure 1: Deposition setup schematics utilizing Ti, Al, and C elemental targets (left) and a Cr–Al–C (2:1:1 composition) composite target (right).*

As-deposited films were annealed in a vacuum furnace with a base pressure below $5\times10^{-4}$ Pa at room temperature and a heating and cooling rate of 10 °C/min. Ti–Al–C coatings were annealed at 800 °C for 4 hours. The Cr–Al–C films were heated to 690 °C and immediately cooled.

Ion-beam-based chemical composition measurements were done by combining time-of-flight elastic recoil detection analysis (ToF-ERDA) and elastic backscattering spectrometry (EBS) at the Tandem Laboratory of Uppsala University [31]. $^{127}I^{8+}$ primary ions at 36 MeV energy were used for ToF-ERDA; the angle between primary ions and the detector telescope was 45°, while the incidence and exit angles were 22.5° with respect to the thin film surface, respectively. Data analysis of time-energy coincidence spectra was carried out with CONTES [32]. 4.5 MeV $^{4}He^{+}$ primary ions were used for EBS and detected at a backscattering angle of 170°. C quantification at the film surfaces was done with the $^{12}C(^{4}He,^{4}He)^{12}C$ elastic resonance at ~4.26 MeV [33], and SIMNRA [34] was employed for data analysis. Average chemical compositions of the thin films reported in the following are obtained from EBS, while ToF-ERDA revealed impurity incorporation in the annealed films.

Structural analyses of the as-deposited and annealed samples were carried out with a Bruker D8 Advance and a Siemens D5000 X-ray diffraction (XRD) system, respectively, both utilizing a Cu Kα radiation source operating at 40 kV and 40 mA. Both systems were aligned in Bragg-Brentano geometry, and the diffractograms were collected with a scan width from 5° to 80°, a step size of 0.5°, and dwell time of 7 s per step. A Bruker AXS D8 Discover General Area Diffraction Detection System was used to elucidate the structural evolution during *in-situ* heating up to 600 °C using an Anton Paar DHS 1100 heating stage. *In-situ* heating was carried out in vacuum, the heating rate was set to 30 °C/min and the diffractograms were recorded in temperature steps of approximately 60 °C, with the incidence angle set to 15 °. A waiting time of 5 minutes was set before each measurement to ensure stable temperature conditions. Diffraction peaks were fitted by a Pearson VII model profile within the TOPAS software; the lattice parameters were calculated using the CellCalc software. The mean crystallite size and strain were computed by the size-strain plot method [35], employing a shape factor of 0.94. The linear coefficient of thermal expansion was determined from the change in lattice parameters during *in-situ* heating.

The microstructure of the annealed films was characterized by scanning transmission electron microscopy (STEM). Cross-sectional lamellae were prepared using focused ion beam techniques in an FEI Helios Nanolab 660 dual-beam microscope. A STEM III detector was employed at 30 kV and 50 pA for acquisition of bright-field micrographs.

Mechanical properties were determined by nanoindentation measurements in a Hysitron TI-900 TriboIndenter, using a Berkovich diamond tip with a 100 nm radius. At least 30 quasistatic indents with maximum loads of 1.5 mN and 3.5 mN in the case of $Ti_3AlC_2$ and $Cr_2AlC$ MAX phase, respectively, were performed, resulting in a contact depth of less than 10 % of the film thickness. The reduced modulus was obtained from the unloading section of load-displacement curves using the method of Oliver and Pharr [36]. The elastic moduli were calculated from the measured reduced moduli using the Poisson's ratio computed from *ab initio* simulations and the isotropic approximation.

**Computational details**

Elastic properties, thermal expansion and the lattice contribution to the thermal conductivity were calculated using density functional theory [37,38], as implemented in the Vienna *ab initio* Simulation Package (VASP, version 5.4.4.) [39,40]. The projected augmented wave (PAW) method [41,42] was employed and electronic exchange and correlation effects were modeled at the generalized gradient approximation level, as parameterized in the well-established PBE (Perdew-Burke-Ernzerhof) functional [43].

The stress-strain method [44] was utilized to calculate the elastic tensor. Six independent universal linear-independent coupling strains were applied, both positive and negative, as described in [44], with the maximum component of every strain being 0.014. Thus, 12 strained configurations in total were calculated, more than would be strictly required on the basis of the crystal symmetry. This was done to minimize the possible numerical errors and add redundancy. The here-reported elastic moduli and Poisson's ratios were obtained from Hill's approximation [45,46]. The hardness was calculated by the revised [47] model of Chen *et al.* [48] as $H_{\mathrm{V}} = 0.92k^{1.137}G^{0.708}$, where *k* is the ratio *G*/*B*, *G* is the shear modulus, and *B* is the bulk modulus. The linear coefficient of thermal expansion was calculated using the quasi-harmonic approximation, implemented in the phonopy code [49]. The lattice part of the thermal conductivity was obtained from the relaxation-time approximation and the solution of the linearized phonon Boltzmann equation, as implemented in phono3py [50,51], post-processing results from anharmonic lattice dynamics calculations performed via VASP. The electronic part of the thermal conductivity over the scattering time ratio was obtained from the Boltzmann transport theory, implemented in the BoltzTRaP2 package [52], with the preceding VASP calculations serving as input data. The value reported herein is the directional average. The scattering time τ was calculated from the electron-phonon interaction in the constant relaxation approximation according to equation (2) [53]:

$$\frac{1}{\tau} = \frac{4\pi k_B T}{\hbar} \int_0^{\infty} \frac{d\omega}{\omega} \frac{x^2}{sinh^2(x)} \alpha_{tr}^2 F(\omega), \qquad (2)$$

where $x = \hbar\omega/(2k_{\mathrm{B}}T)$, $T$ is the temperature, $k_{\mathrm{B}}$ is the Boltzmann constant, and $\omega$ is the phonon frequency. The transport Eliashberg spectral function $\alpha_{tr}^2F(\omega)$ was calculated by the EPW package [54], on top of the density functional perturbation theory (DFPT) phonon calculations performed using the QuantumEspresso (QE) DFT suite [55–57].

Spin polarization was considered for $Cr_2AlC$ and modeled with the most stable antiferromagnetic ordering reported in the literature [58]; however, for the QE DFPT and EPW calculations, a non-spin polarized setup was used. Additionally, a Hubbard *U* correction of 2 eV for the Cr 3*d* states was utilized [59], yielding bulk moduli and lattice parameters closest to the experimental values (see Supplementary Material S1). Specific numerical settings include a Monkhorst-Pack grid [60], a plane-wave cutoff energy of 500 eV and *k*-point grids with a density of at least 800 *k*-points Å$^3$ for the VASP calculations, except for lattice dynamics which were performed with a single *k*-point at the Γ point. The force constants of the 2$^{nd}$ and 3$^{rd}$ order were calculated with 2×2×2 supercells employing a cut-off pair distance of 5 Å. The employed values should guarantee an energy convergence better than ~ 1.5 meV per atom, force convergence better than ~ 2 meV/Å and stress convergence better than 5 % in terms of differences between steps (see Supplementary Material S2). For the lattice thermal conductivity, the reciprocal spaces were sampled with a 19×19×19 mesh. QE-based simulations utilized the recommended pseudopotentials from the “A standard solid-state pseudopotential” (SSSP) library v1.1.2 precision set [61,62], a kinetic energy cut-off of 750 eV for the wavefunctions, Monkhorst-Pack *k*-point grids with a density of at least 600 *k*-points Å$^3$ and a *q*-point density of at least 80 *q*-points Å$^3$. For the EPW calculations, the QE electron and phonon band structure was interpolated to 8 times denser *k* and *q* grids using the maximally localized Wannier functions [63]. The phonopy calculations used supercell sizes of 240 and 360 atoms for $Cr_2AlC$ and $Ti_3AlC_2$, respectively, to exclude self-interactions of displaced atoms with their periodic images. All calculation results with further details are available in the NOMAD archive [64].

## Results and discussion

### Chemical composition and structure

The chemical composition of as-deposited and vacuum-annealed MAX phase coatings is provided in Table 1. The average concentrations of MAX phase constituents were determined by EBS, while the depth-dependent oxygen incorporation was characterized by ToF-ERDA depth profiling. The chemical compositions of both MAX phase thin films were consistent with a 2:1:1 (further expressed as 211) stoichiometric ratio. Regardless of the average 211 composition, the titanium-containing MAX phase formed $Ti_3AlC_2$. Note that the labels of annealed samples are based on their crystal structures. These are stated at this point and further discussed in the section regarding XRD (see below) and the Supplementary Material S3. The chemical depth profiles measured by ToF-ERDA are shown in Figure 2. Oxygen was preferentially incorporated in the region near the surface for both Ti–Al–C and Cr–Al–C coatings, which can be understood in the as-deposited state due to atmosphere exposure [65]. For the annealed samples, it may be linked to an increase in base pressure during vacuum annealing, while the larger oxygen uptake in $Ti_3AlC_2$ might originate from microstructural differences (see below).

| | Ti (at. %) | Cr (at. %) | Al (at. %) | C (at. %) | O (at. %) |
|---|---|---|---|---|---|
| Ti–Al–C as-deposited | 46.4 ± 1.9 | | 27.0 ± 1.1 | 26.5 ± 1.1 | 0.1 ± 0.1 |
| $Ti_3AlC_2$ annealed | 46.9 ± 1.9 | | 27.2 ± 1.1 | 25.3 ± 1.0 | 0.6 ± 0.1 |
| Cr–Al–C as-deposited | | 48.5 ± 1.9 | 26.0 ± 1.0 | 25.3 ± 1.0 | 0.3 ± 0.1 |
| $Cr_2AlC$ annealed | | 47.6 ± 1.9 | 26.0 ± 1.0 | 22.6 ± 0.9 | 3.8 ± 0.2 |

*Table 1: Chemical compositions of the as-deposited and annealed coatings as measured by ToF-ERDA and EBS. Average oxygen concentrations are obtained by excluding the oxygen-rich surface regions.*

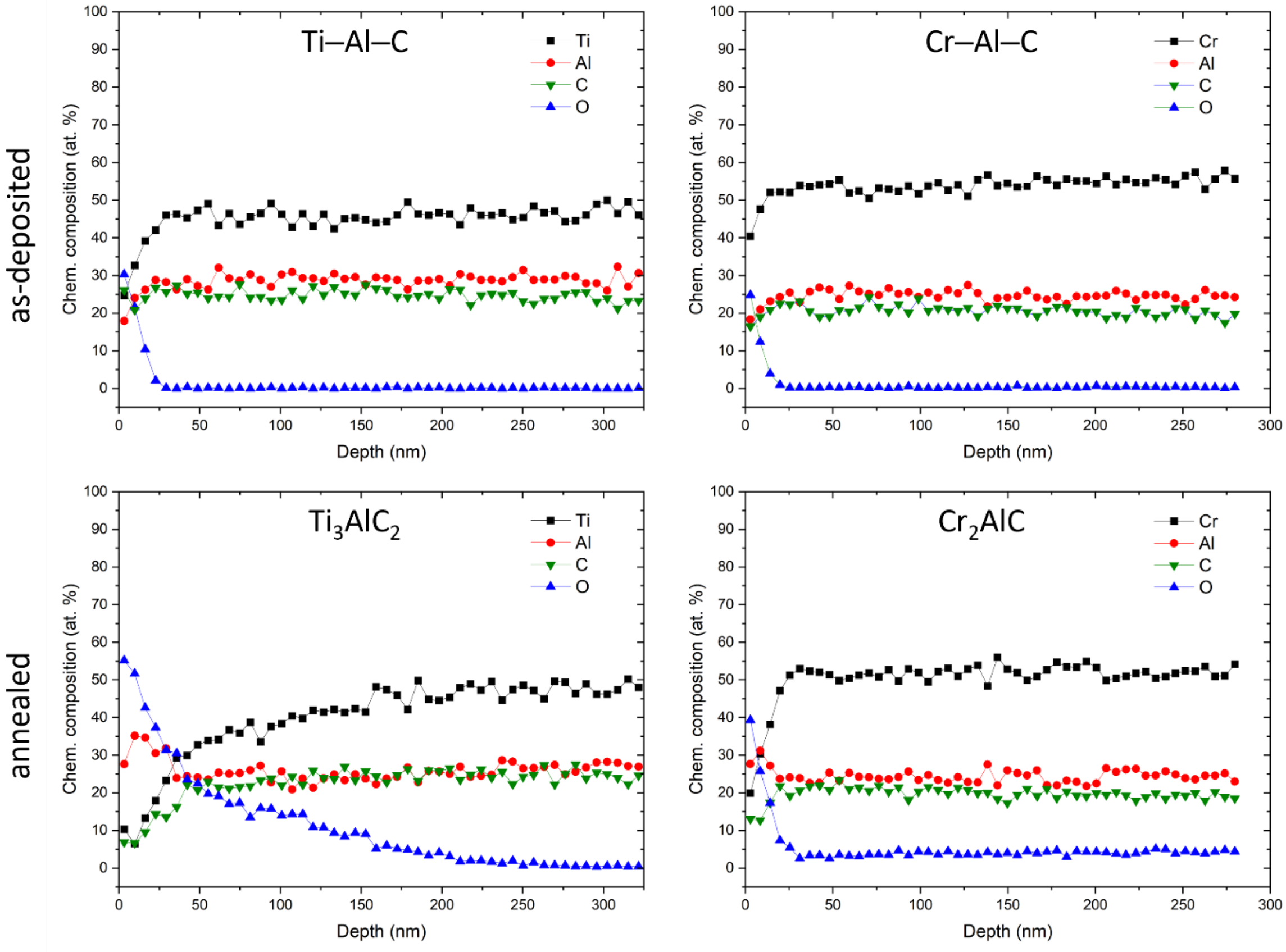

*Figure 2: Concentration depth profiles of as-deposited Ti–Al–C and Cr–Al–C films, and annealed $Ti_3AlC_2$ and $Cr_2AlC$ films measured by ToF-ERDA. The depth was calculated*

*assuming a $Ti_3AlC_2$ density of 4.20 g·cm$^{-3}$ [66] and a $Cr_2AlC$ density of 5.17 g·cm$^{-3}$ [67].*

The XRD patterns of the as-deposited films and annealed MAX phases are shown in Figure 3. The diffraction peak at approximately 43 ° originates from the MgO substrate. The other peaks correspond to the MAX phases. The formation of the $Ti_3AlC_2$ MAX phase is verified by a combination of differential scanning calorimetry and X-ray diffraction (see Supplementary Material S3). Computed lattice parameters and unit cell volumes of annealed coatings are in good agreement with the values obtained from DFT and stated in powder diffraction databases, as shown in Table 2. The mean crystallite sizes of 19 and 88 nm and strains of $9.9\times10^{-4}$ and $5.2\times10^{-4}$ were computed from the XRD patterns of $Ti_3AlC_2$ and $Cr_2AlC$, respectively. The smaller grain size of $Ti_3AlC_2$ compared to that of $Cr_2AlC$ can be seen from the cross-sectional STEM images shown in Figure 4. The larger fraction of grain boundaries in the annealed $Ti_3AlC_2$ compared to $Cr_2AlC$ is most likely the reason for the differences in oxygen incorporation from residual gas during annealing.

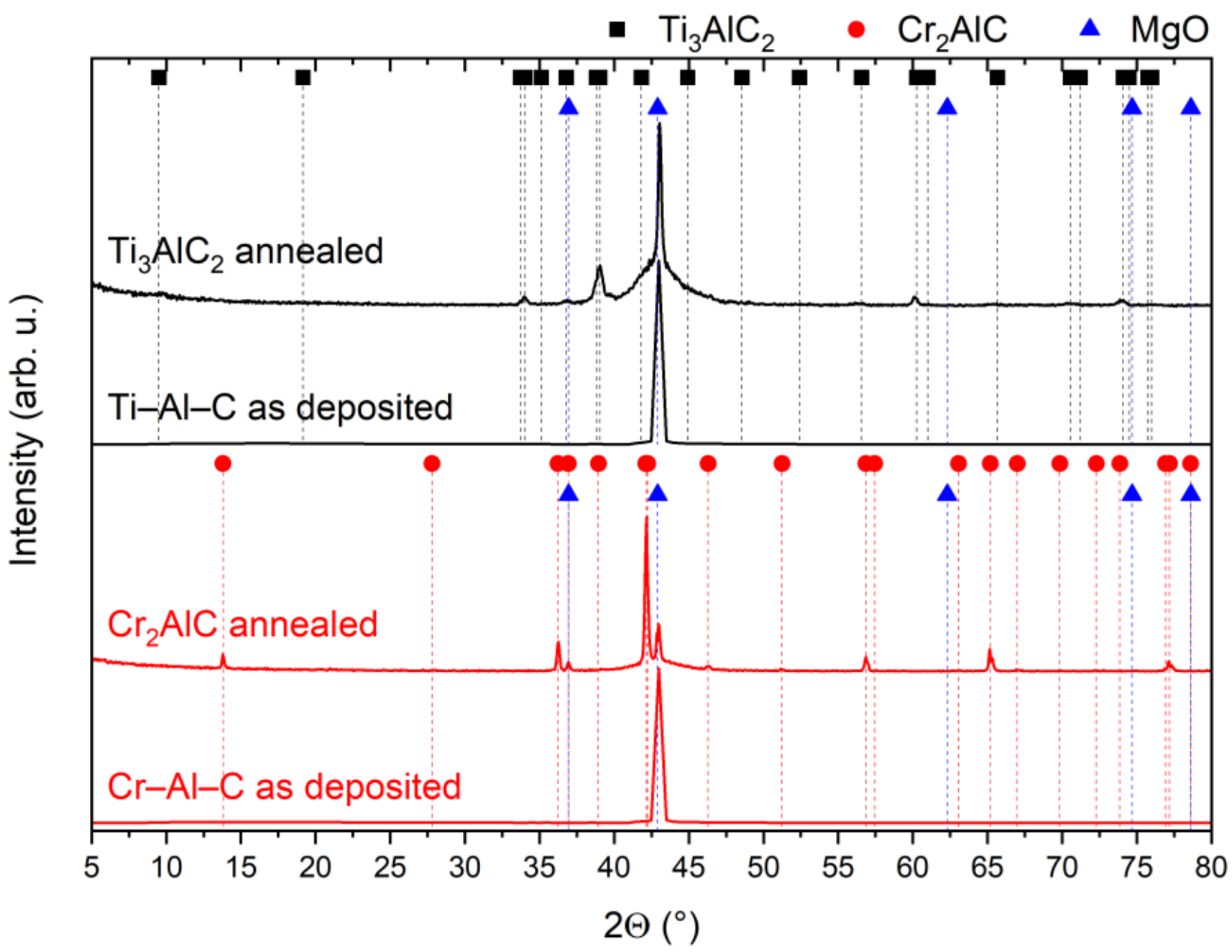

*Figure 3: Diffraction patterns of as-deposited / annealed Ti–Al–C / $Ti_3AlC_2$ (black) and Cr–Al–C / $Cr_2AlC$ (red) films.*

| | | $Ti_3AlC_2$ | $Cr_2AlC$ |
|---|---|---|---|
| *a* (Å) | Experiment | 3.07 ±0.01 | 2.86 ± 0.01 |
| | Simulation | 3.08 | 2.88 |
| | Database | 3.069 | 2.86 |
| *c* (Å) | Experiment | 18.57 ± 0.02 | 12.83 ± 0.01 |
| | Simulation | 18.66 | 12.87 |
| | Database | 18.501 | 12.82 |
| *V* ($Å^3$) | Experiment | 151.57 ± 0.01 | 90.88 ± 0.01 |
| | Simulation | 153.30 | 92.44 |
| | Database | 150.91 | 90.81 |

*Table 2: Lattice parameters and unit cell volumes obtained from the measured XRD patterns and the DFT simulations and extracted from a database (PDF 00-052-0875 and PDF 01-089-2275 for $Ti_3AlC_2$ and $Cr_2AlC$, respectively).*

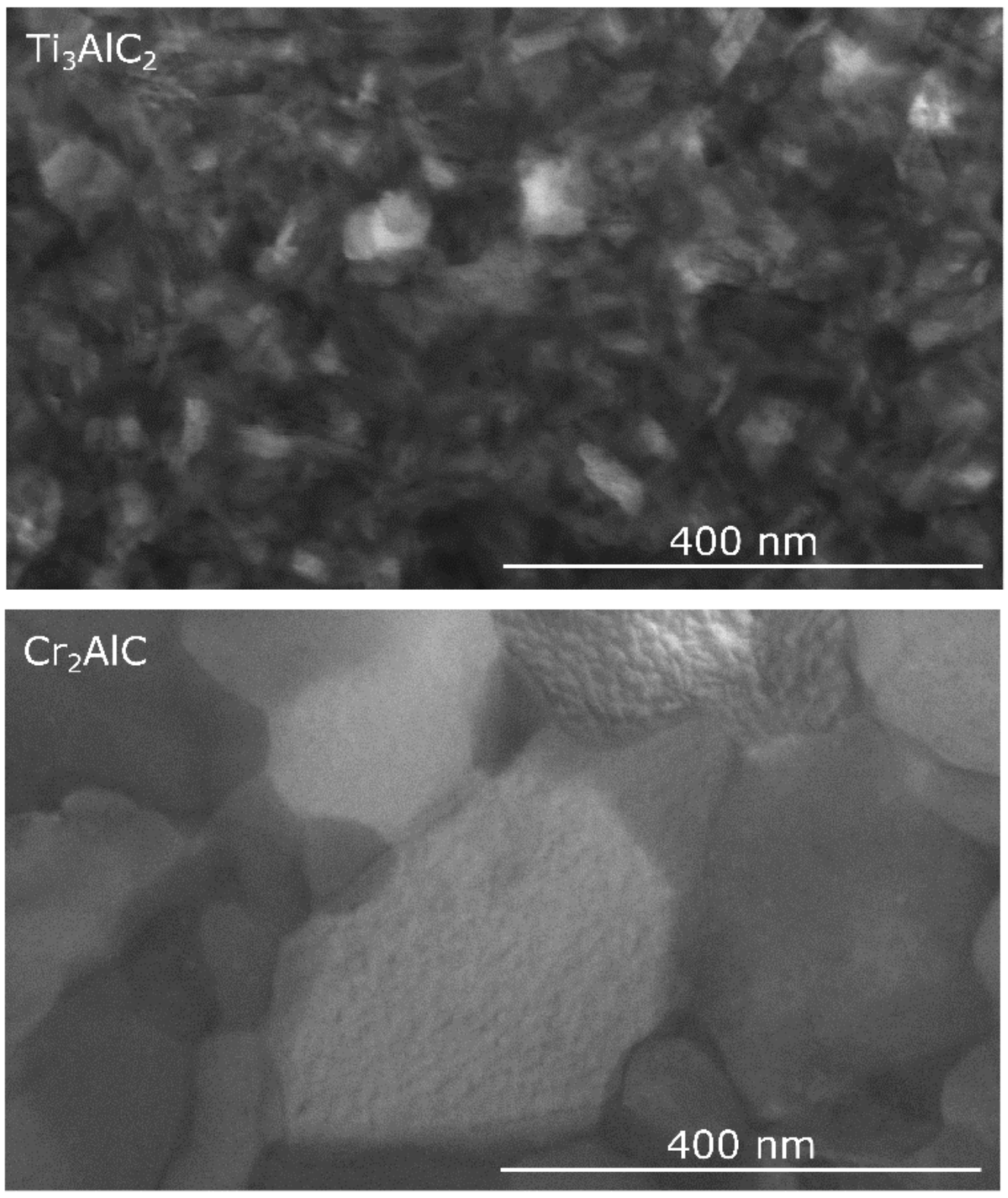

*Figure 4: Bright-field STEM images of annealed $Ti_3AlC_2$ (top) and $Cr_2AlC$ (bottom) MAX phase films.*

**Elastic properties**

The hardness *H* and elastic modulus *E* of the annealed coatings are given in Table 3 as obtained from nanoindentation measurements and are compared to simulations. It should be noted that these measurements were affected by the oxide scale formed on the surface of the coating. Nevertheless, the measured hardness was 18.1 ± 1.7 GPa and 13.3 ± 0.7 GPa for $Ti_3AlC_2$ and $Cr_2AlC$, respectively, with both values close to those reported for the thin film MAX phases (20.7 GPa for $Ti_3AlC_2$ [68] and 13 ± 2 GPa [69] and 15 ± 0.4 GPa [70] for $Cr_2AlC$). The hardness measured for thin film samples is often larger compared to the corresponding bulk samples as the thin film grain sizes are often smaller than in bulk [69], which is consistent with the Hall-Petch relationship. The measured hardness was lower compared to the quantum-mechanically predicted 23 GPa for $Ti_3AlC_2$ and 21 GPa for $Cr_2AlC$. Both the measured and computed hardness values were higher for $Ti_3AlC_2$ vs. $Cr_2AlC$ when comparing the MAX phases to each other. The particularly large difference between measured and computed *H* for $Cr_2AlC$ partially stems from the high sensitivity of the individual elastic tensor components from DFT simulations, from which the hardness is extracted, towards both magnetic ordering and the choice of the *U* parameter used for the Hubbard correction [58,59]. As such, even small changes in these values may induce large differences in *H*. Generally speaking, the dependence upon the elastic tensor components leads to deviations between approx. 10 and 20 % of the DFT-derived values for *E* vs. the experimental values, with the error propagating onto the hardness [71].

While the hardness is not necessary for the estimation of the thermal shock parameter, it can be used to estimate the flexural strength, which is required per equation (1). Assuming that the flexural strength $\sigma_f$ is equal to the tensile strength $\sigma$, it can be generally obtained via the relationship $3\sigma \approx H$ [72–74]. Zhang *et al.* [72] studied the relationship between the strength and hardness of work-hardened metals, bulk metallic glasses (BMG), and ceramics. It was concluded that $3\sigma \approx H$ holds for work-hardened crystalline materials and the shearable BMGs, but the factor is significantly larger than 3 in the brittle BMGs and ceramics. This is confirmed by several experimental data sets of different ceramics for example $Si_3N_4$ ($H/\sigma_f \approx 51$), $ZrO_2$ ($H/\sigma_f \approx 122$), and TiC ($H/\sigma_f \approx 235$) [75].

Thus, to improve the precision of this estimate, the ratio between hardness and flexural strength values stated in the literature is utilized as a denominator in this study. Tzenov *et al.* [76] reported 3.5 GPa hardness and 375 MPa flexural strength of bulk $Ti_3AlC_2$, resulting in a ratio of 9.33. For $Cr_2AlC$, the ratio is 10.45, obtained from hardness and flexural strength of 4.9 GPa and 469 MPa, respectively, as stated by Ying *et al.* [77]. Hence, we utilize these ratios for the here investigated coatings to estimate flexural strength data based on measured coating hardness values. The estimated experimental and quantum-mechanically predicted $Ti_3AlC_2$ flexural strengths are 1.94 ± 0.18 GPa and 2.45 GPa, respectively. In the case of $Cr_2AlC$, the estimated experimental and predicted flexural strengths are 1.27 ± 0.07 GPa and 2.01 GPa, respectively.

The experimentally obtained elastic modulus of $Ti_3AlC_2$ of 335 ± 21 GPa is in good agreement with a computed value of 305 GPa, in line with the 304 GPa reported by Wan *et al.* [78]. A larger discrepancy is observed for $Cr_2AlC$, with measured and simulated elastic modulus values of 341 ± 9 GPa and 287 GPa, respectively. Besides the sensitivity of the elastic constants vs. the magnetic model and the *U* parameter discussed above, and the corresponding inherent uncertainty of the simulated value, the stress state of the deposited coating may also contribute to the larger experimentally measured value of *E* [79]. Nevertheless, these values are in semiquantitative agreement with previously reported experimental values of 279 GPa [80], 282 GPa [77], 298 GPa [69], and simulated values of 328.62 GPa [81], and 357.7 GPa [82]. The significant difference between the experimental and simulated moduli in the literature data, as well as the larger scattering among the latter, is yet again likely caused by differing magnetic ordering models and *U* parameters.

Within the isotropic approximation, Poisson's ratios of 0.179 and 0.188 were used to determine the elastic modulus of $Ti_3AlC_2$ and $Cr_2AlC$, respectively, in line with Poisson's ratios of 0.2 [66] and 0.153 [83] as given in literature.

| | Nanoindentation | | DFT prediction | | Ref. |
|---|---|---|---|---|---|
| | *H* (GPa) | *E* (GPa) | *H* (GPa) | *E* (GPa) | *H* (GPa) |
| $Ti_3AlC_2$ | 18.1 ± 1.7 | 335 ± 21 | 23 | 305 | 20.7 [68] |
| $Cr_2AlC$ | 13.3 ± 0.7 | 341 ± 9 | 21 | 287 | 15 ± 0.4 [70] |

*Table 3: Hardness and elastic modulus of annealed $Ti_3AlC_2$ and $Cr_2AlC$ MAX phase films determined both by nanoindentation and DFT calculation and cited thin film hardness.*

## Thermal properties

Figure 5 shows the computed evolutions of electronic, phononic, and total thermal conductivities as a function of temperature, with all components decreasing within the temperature range of 300 to 1000 K for both phases. The electronic thermal conductivity is the dominant contribution to the total thermal conductivity. The larger scattering of electrons on phonons causes a decrease in thermal conductivity at higher temperatures. The computed $Ti_3AlC_2$ thermal conductivity of 33.8 $W{\cdot}m^{-1}{\cdot}K^{-1}$ at 300 K agrees with the reported thermal conductivity of 40 $W{\cdot}m^{-1}{\cdot}K^{-1}$ at room temperature [84]. In the case of $Cr_2AlC$, the computed thermal conductivity is 32.4 $W{\cdot}m^{-1}{\cdot}K^{-1}$ at room temperature. This is in line with the directionally averaged thermal conductivity of 34.73 $W{\cdot}m^{-1}{\cdot}K^{-1}$ computed based on the measurements of Champagne *et al.* [85] of the in-plane thermal conductivity of 41.3 $W{\cdot}m^{-1}{\cdot}K^{-1}$ and out-of-plane thermal conductivity of 21.6 $W{\cdot}m^{-1}{\cdot}K^{-1}$ in a single crystal $Cr_2AlC$ at room temperature.

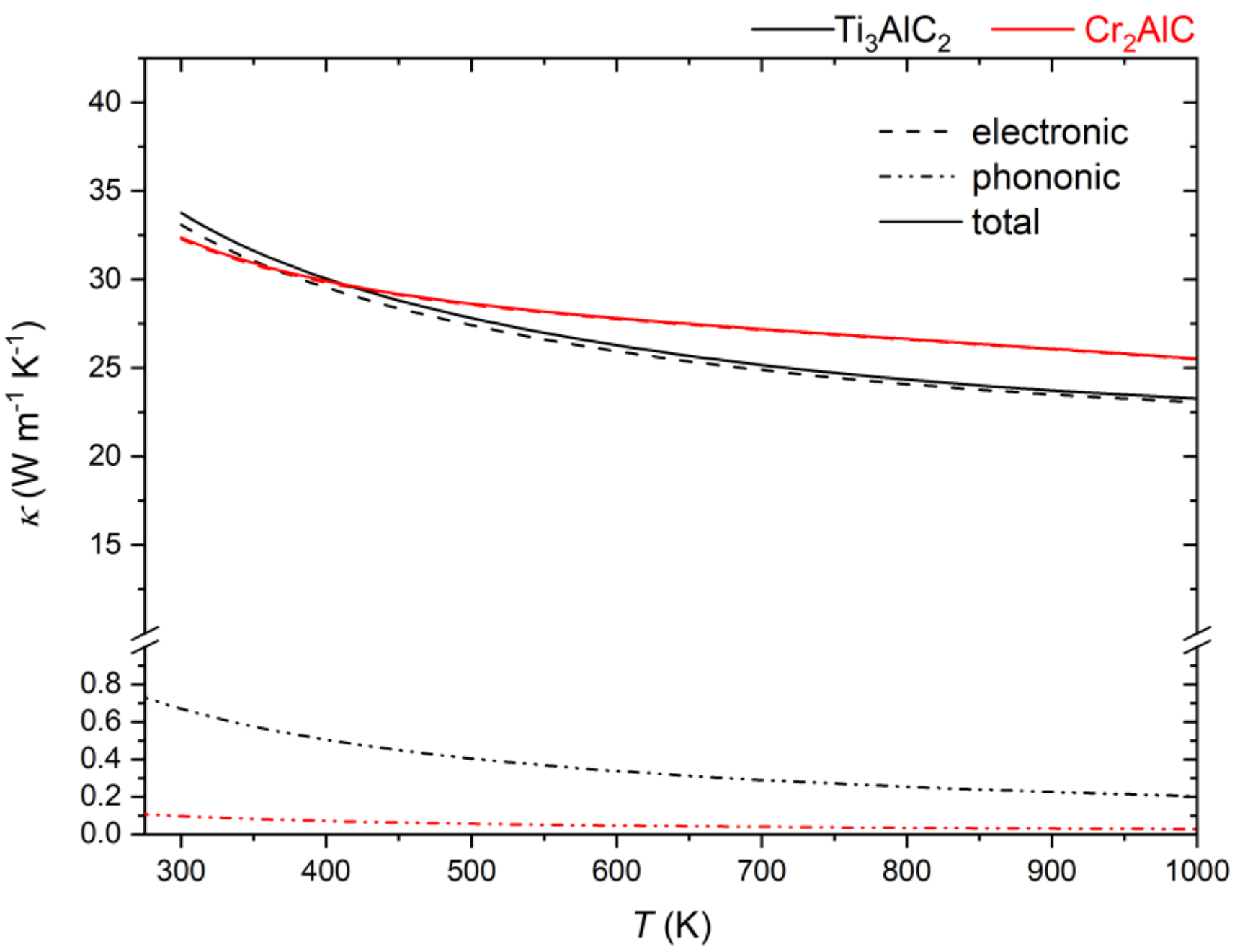

*Figure 5: Electronic, phononic, and total thermal conductivities as a function of temperature as determined by DFT calculations.*

The experimental and computed average linear coefficient of thermal expansion is shown in Figure 6. Experimentally, average LCTE was determined from the change of lattice parameters during *in-situ* heating between 300 and 873 K. The quantum-mechanically predicted average LCTE increases with increasing temperature, consistent with the findings of Ying *et al.* [77]. The measured LCTE of $(7.3 \pm 1.4) \times 10^{-6}$ $K^{-1}$ and $(11.8 \pm 0.4) \times 10^{-6}$ $K^{-1}$ of $Ti_3AlC_2$ and $Cr_2AlC$, respectively, are in good agreement with literature data ($9.0 \times 10^{-6}$ $K^{-1}$ for $Ti_3AlC_2$ [76] and $12.5 \times 10^{-6}$ $K^{-1}$ for $Cr_2AlC$ [77]). The mismatch between calculated and experimentally determined LCTE for $Cr_2AlC$ could be caused by difficulties to properly describe the magnetic state of the system at higher temperatures, as extremely computationally expensive non-collinear calculations would need to be performed to correctly reflect paramagnetic behavior, which is likely present in $Cr_2AlC$ at higher temperatures.

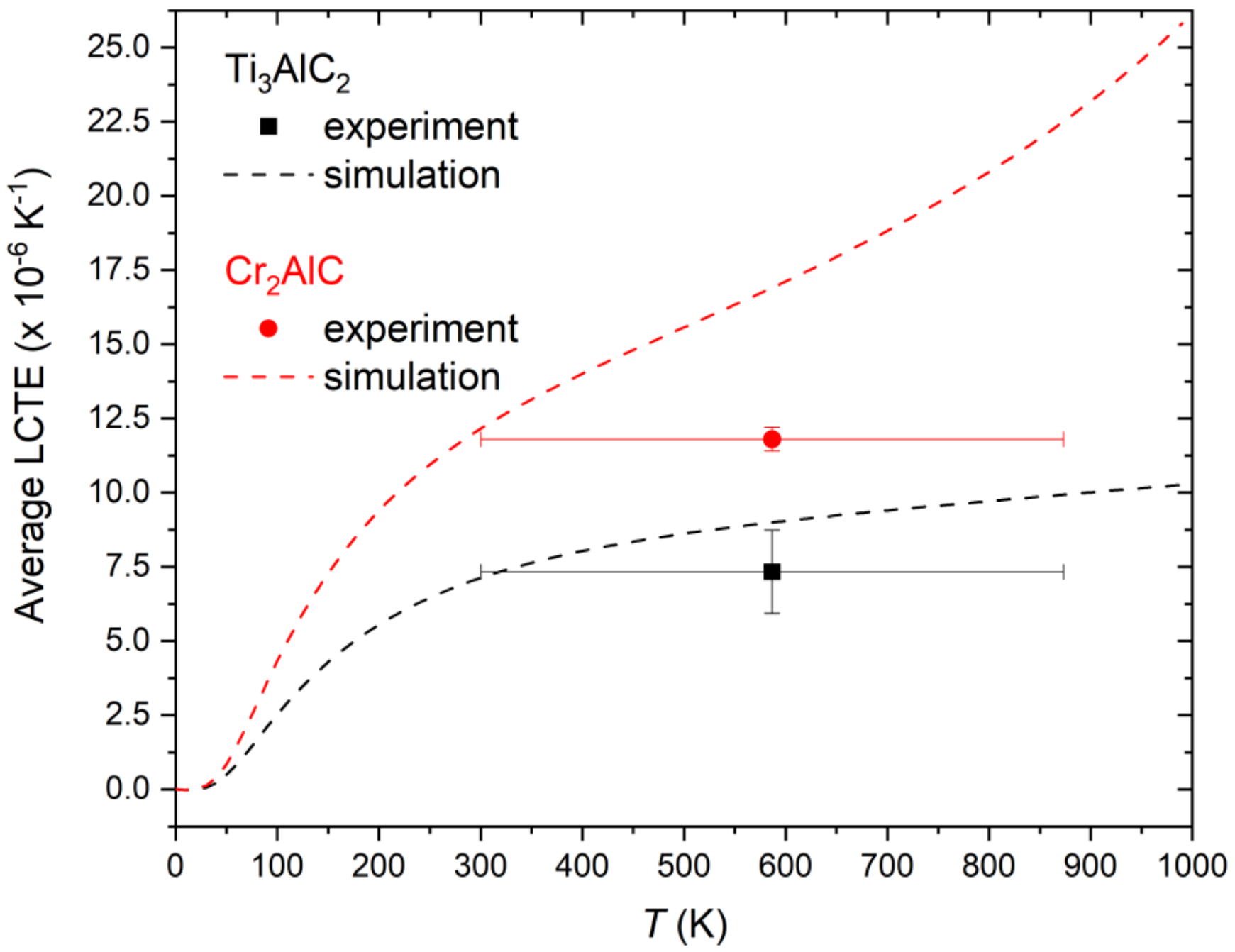

*Figure 6: Average linear coefficient of thermal expansion of annealed $Ti_3AlC_2$ and $Cr_2AlC$ MAX phase films determined both experimentally (symbol) and via DFT calculations (dashed line).*

Table 4 summarizes all properties needed for the calculation of the thermal shock parameter $R_T$ and their deviations, with the intent to compare experiments and predictions. Literature data are used for the experimental thermal conductivity [84,85]. Furthermore, the Poisson's ratios considered for the experimental assessment are computed, the quantum-mechanically predicted LCTE is averaged across the temperature range between 300 and 870 K, and all the reference values were measured utilizing bulk samples. We thus stress that the calculated $R_T$ values, in light of these approximations, serve well for a semiquantitative comparison of both material systems while not necessarily being quantitatively exact.

The experimentally derived thermal shock parameter of the $Ti_3AlC_2$ MAX phase is 26.05 ± 5.81 kW·m$^{-1}$. The uncertainty $U_{\mathrm{Exp,\ RT}}$ ≈ 22 % is primarily induced by the contribution of the average LCTE uncertainty, $U_{\mathrm{Exp},LCTE}$ ≈ 19 %. The value obtained from the DFT simulations is 25.05 kW m$^{-1}$, with the difference between the experimental mean and the value from simulations being $D_{\mathrm{Exp\text{-}Calc}}$ ≈ 4 %, in very good agreement with the experimental dataset. As for the individual contributions to $R_T$, $D_{Exp-Calc}$ is comparable for flexural strength, thermal conductivity and the average LCTE at around 20 %, with a deviation of 10 % for the elastic modulus. Thus, the errors of all parameters affect the result in a comparable way for $Ti_3AlC_2$.

In the case of $Cr_2AlC$, the experimentally derived $R_T$ = 8.90 ± 0.62 kW m$^{-1}$, with the highest uncertainty contribution stemming from the flexural strength term with $U_{Exp,\sigma f} \approx$ 6 % and the overall uncertainty being $U_{Exp,\ RT} \approx$ 7 %. Thus, the scattering of the experimentally obtained values is lower for all examined parameters as compared to $Ti_3AlC_2$. Comparing the thermal shock parameter obtained from experimental data and the DFT predictions for $Cr_2AlC$, $D_{Exp-Calc} \approx$ 17.9 %. The major contributions to this are the flexural strength, where $D_{Exp-Calc} \approx$ 36.8 % and the average LCTE, where $D_{Exp-Calc} \approx$ 30.6 %. These contributions to $R_T$ are significantly overestimated in the simulations, whereas $\kappa$ and $E$ are underestimated compared to the experimental measurements. The discrepancy $D_{Exp-Calc}$ is likely caused by challenge to model the experimentally relevant magnetic state accurately, as discussed in the section on hardness, elastic modulus, and LCTE. Generally speaking, $Ti_3AlC_2$ exhibits a higher $R_T$ in comparison to $Cr_2AlC$ for both experiment- and simulation-derived datasets, primarily driven by the difference in *LCTE*.

When calculating $R_T$ using values reported in literature and comparing it to the presented results, the striking difference is caused mainly by the larger flexural strength due to the higher hardness obtained in this study. The thermal shock parameter of the $Ti_3AlC_2$ MAX phase calculated from the averaged literature values is 4.20 ± 0.41 kW·m$^{-1}$, with the highest deviation pertaining to the flexural strength at $D_{Ref.,\sigma f} \approx$ 8 %. Correspondingly, the computed thermal shock parameter of $Cr_2AlC$ using averaged literature values results in 3.60 ± 0.68 kW·m$^{-1}$. For this phase, the reported values deviate mostly in Poisson's ratio with $D_{Ref.,\nu} \approx 19$ %; the second largest deviation occurs for the flexural strength with $D_{Ref.,\sigma f} \approx 17$ %.

| **(a) $Ti_3AlC_2$** | Exp. | Calc. | Ref. | $U_{Exp}$ (%) | $D_{Exp\text{-}Calc}$ (%) | $D_{Ref}$ (%) |
|---|---|---|---|---|---|---|
| $\sigma_f$ (GPa) | 1.94 ± 0.18 | 2.45 | 0.3203 [78]<br>0.340 [86]<br>0.375 [76] | 9.3 | 20.8 | 8.0 |
| $\kappa_{300\ K}$ (W m$^{-1}$ K$^{-1}$) | 40 [84] | 33.8 | 40 [84] | | 18.3 | |
| $\nu$ | 0.179 | 0.179 | 0.2 [66] | | | |
| *average LCTE* ($\times10^{-6}$ K$^{-1}$) | 7.3 ± 1.4 | 8.9 | 8.1 [87]<br>9.0 [76]<br>9.0 [13]<br>9.0 [87]<br>9.2 [88] | 19.2 | 18.0 | 4.9 |
| $E$ (GPa) | 335 ± 22 | 305 | 288 [89]<br>297.5 [66]<br>304 [78] | 6.6 | 9.8 | 2.7 |
| $R_T$ (kW m$^{-1}$) | 26.05 ± 5.81 | 25.05 | 4.20 ± 0.41 | 22.3 | 4.0 | 9.8 |

| **(b) $Cr_2AlC$** | Exp. | Calc. | Ref. | $U_{Exp}$ (%) | $D_{Exp\text{-}Calc}$ (%) | $D_{Ref}$ (%) |
|---|---|---|---|---|---|---|
| $\sigma_f$ (GPa) | 1.27 ± 0.07 | 2.01 | 0.305 [90]<br>0.378 [83]<br>0.469 [77]<br>0.483 [67]<br>0.483 [91]<br>0.498 [92]<br>0.513 [90] | 5.5 | 36.8 | 17.1 |
| $\kappa_{300\ K}$ (W m$^{-1}$ K$^{-1}$) | 34.73 [85] | 32.4 | $\kappa_{\parallel}$ = 41.3<br>$\kappa_{\perp}$ = 21.6<br>$\kappa_{av}$ = 34.73 [85] | | 7.2 | |
| $\nu$ | 0.188 | 0.188 | 0.153 [83]<br>0.2 [26] | | | 18.8 |
| *average LCTE* ($\times10^{-6}$ K$^{-1}$) | 11.8 ± 0.4 | 17.0 | 12.5 [77]<br>12.6 [93]<br>12.8 [94]<br>13.1 [92]<br>13.3 [83] | 3.4 | 30.6 | 2.6 |
| $E$ (GPa) | 341 ± 9 | 287 | 245 [26]<br>278 [83]<br>282 [77]<br>288 [91] | 2.6 | 18.8 | 6.6 |
| | | | 289 [67] | | | |
| $R_T$ (kW m$^{-1}$) | 8.90 ± 0.62 | 10.84 | 3.60 ± 0.68 | 7.0 | 17.9 | 18.9 |

*Table 4: Experimentally obtained, quantum-mechanically predicted, and cited thermomechanical properties of $Ti_3AlC_2$ (a) and $Cr_2AlC$ (b) MAX phases, and the calculated thermal shock parameter $R_T$ for all datasets. The uncertainties of the experimental values $U_{exp}$ and the experiment-theory difference $D_{Exp\text{-}Cal}$ are given. Note that the reference thermomechanical properties cited from literature were measured utilizing bulk samples. The reference deviations $D_{Ref}$ are calculated based on the average and corresponding standard deviation of the listed published values.*

Correlating the observations for MAX phase coatings, both measurements and simulations indicate a superior thermal shock behavior of $Ti_3AlC_2$ in comparison with $Cr_2AlC$, induced primarily by the higher linear coefficient of thermal expansion of $Cr_2AlC$. The results presented herein underline the capacity of *ab initio* calculations for a semiquantitative estimation of the thermal shock parameter of MAX phases.

## Conclusion

In this study, the thermal shock behavior of $Ti_3AlC_2$ and $Cr_2AlC$ MAX phase thin films is investigated. The resistance to thermal shock can be described by the thermal shock parameter ($R_T$), which is proportional to the thermal conductivity, flexural strength, and Poisson’s ratio and inversely proportional to the linear coefficient of thermal expansion and elastic modulus. These thermomechanical properties were investigated by both experiments and DFT simulations. The $R_T$ of both $Ti_3AlC_2$ and $Cr_2AlC$ obtained through quantum-mechanical predictions is in very good agreement with the experimentally obtained $R_T$. The larger discrepancy in the case of $Cr_2AlC$ is likely caused by the difficulty to correctly model the experimentally relevant magnetic state. Upon comparison of the MAX phases, both measurements and quantum-mechanical predictions indicate a superior thermal shock resistance for $Ti_3AlC_2$ vs. $Cr_2AlC$, driven mainly by the higher linear coefficient of thermal expansion of $Cr_2AlC$. The results presented herein underline the potential of *ab initio* calculations for a semiquantitative evaluation of the thermal shock behavior of ionically-covalently bonded materials.

## Acknowledgments

This research was funded by the German Research Foundation (DFG, 2866/4-1 and 735/38-1). Transnational ion beam analysis access has been supported by the RADIATE project under Grant Agreement 824096 from the EU Research and Innovation programme HORIZON 2020. Accelerator operation at Uppsala University has been supported by the Swedish research council VR-RFI under grant agreement #2019-00191. We also gratefully acknowledge the provision of computational resources by the IT Center of RWTH Aachen University within the framework of the Jülich-Aachen Research Alliance (JARA) jara0131 and jara0151 computing time grants. PO acknowledges the support by the project LM2023039 funded by the Ministry of Education, Youth and Sports of the Czech Republic. We thank the Chair of Ceramics at the Institute of Mineral Engineering for access to the X-ray diffraction systems.

## References

[1] T.J. Lu, N.A. Fleck, The thermal shock resistance of solids, Acta Materialia 46 (1998) 4755–4768. https://doi.org/10.1016/S1359-6454(98)00127-X.

[2] D.R. Askeland, P.P. Phulé, W. Wright, The science and engineering of materials, sixthth ed., Cengage Learning, Stamford, op. 2011.

[3] D.P.H. HASSELMAN, Unified Theory of Thermal Shock Fracture Initiation and Crack Propagation in Brittle Ceramics, Journal of the American Ceramic Society 52 (1969) 600–604. https://doi.org/10.1111/j.1151-2916.1969.tb15848.x.

[4] W.F. Krupke, M.D. Shinn, J.E. Marion, J.A. Caird, S.E. Stokowski, Spectroscopic, optical, and thermomechanical properties of neodymium- and chromium-doped gadolinium scandium gallium garnet, J. Opt. Soc. Am. B 3 (1986) 102. https://doi.org/10.1364/JOSAB.3.000102.

[5] D. Music, P. Bliem, M. Hans, Holistic quantum design of thermoelectric niobium oxynitride, Solid State Communications 212 (2015) 5–9. https://doi.org/10.1016/j.ssc.2015.03.020.

[6] T. El-Raghy, M.W. Barsoum, A. Zavaliangos, S.R. Kalidindi, Processing and Mechanical Properties of $Ti_3SiC_2$: II, Effect of Grain Size and Deformation Temperature, Journal of the American Ceramic Society 82 (1999) 2855–2860. https://doi.org/10.1111/j.1151-2916.1999.tb02167.x.

[7] D. Music, O. Kremer, G. Pernot, J.M. Schneider, Designing low thermal conductivity of $RuO_2$ for thermoelectric applications, Appl. Phys. Lett. 106 (2015) 63906. https://doi.org/10.1063/1.4909513.

[8] R. Daniel, D. Holec, M. Bartosik, J. Keckes, C. Mitterer, Size effect of thermal expansion and thermal/intrinsic stresses in nanostructured thin films: Experiment and model, Acta Materialia 59 (2011) 6631–6645. https://doi.org/10.1016/j.actamat.2011.07.018.

[9] M. Ahlgren, H. Blomqvist, Influence of bias variation on residual stress and texture in TiAlN PVD coatings, Surface and Coatings Technology 200 (2005) 157–160. https://doi.org/10.1016/j.surfcoat.2005.02.078.

[10] J.C. Han, Thermal shock resistance of ceramic coatings, Acta Materialia 55 (2007) 3573–3581. https://doi.org/10.1016/j.actamat.2007.02.007.

[11] B. Liang, C. Ding, Thermal shock resistances of nanostructured and conventional zirconia coatings deposited by atmospheric plasma spraying, Surface and

Coatings Technology 197 (2005) 185–192. https://doi.org/10.1016/j.surfcoat.2004.08.225.

[12] K. Kokini, J. DeJonge, S. Rangaraj, B. Beardsley, Thermal shock of functionally graded thermal barrier coatings with similar thermal resistance, Surface and Coatings Technology 154 (2002) 223–231. https://doi.org/10.1016/S0257-8972(02)00031-2.

[13] M.W. Barsoum, The $M_{N+1}AX_N$ phases: A new class of solids, Progress in Solid State Chemistry 28 (2000) 201–281. https://doi.org/10.1016/S0079-6786(00)00006-6.

[14] M.W. Barsoum, T. El-Raghy, Synthesis and Characterization of a Remarkable Ceramic: $Ti_3SiC_2$, Journal of the American Ceramic Society 79 (1996) 1953–1956. https://doi.org/10.1111/j.1151-2916.1996.tb08018.x.

[15] M.W. Barsoum, M. Radovic, Elastic and Mechanical Properties of the MAX Phases, Annu. Rev. Mater. Res. 41 (2011) 195–227. https://doi.org/10.1146/annurev-matsci-062910-100448.

[16] M.W. Barsoum, MAX phases: Properties of machinable ternary carbides and nitrides, Wiley-VCH, Weinheim, 2013.

[17] Y. Bai, F. Kong, X. He, N. Li, X. Qi, Y. Zheng, C. Zhu, R. Wang, A.I. Duff, Thermal shock behavior of $Ti_2AlC$ from 200°C to 1400°C, Journal of the American Ceramic Society 100 (2017) 4190–4198. https://doi.org/10.1111/jace.14965.

[18] Y.W. Bao, X.H. Wang, H.B. Zhang, Y.C. Zhou, Thermal shock behavior of $Ti_3AlC_2$ from between 200°C and 1300°C, Journal of the European Ceramic Society 25 (2005) 3367–3374. https://doi.org/10.1016/j.jeurceramsoc.2004.08.026.

[19] H.B. Zhang, Y.C. Zhou, Y.W. Bao, M.S. Li, Abnormal thermal shock behavior of $Ti_3SiC_2$ and $Ti_3AlC_2$, J. Mater. Res. 21 (2006) 2401–2407. https://doi.org/10.1557/jmr.2006.0289.

[20] W. Ding, B. Hu, S. Fu, D. Wan, Y. Bao, Q. Feng, S. Grasso, C. Hu, Ultra-Fast Thermal Shock Evaluation of $Ti_2AlC$ Ceramic, Materials (Basel) 15 (2022). https://doi.org/10.3390/ma15196877.

[21] B. Hu, Y. Bao, X. Su, D. Wan, Q. Feng, S. Grasso, C. Hu, Comparative investigation of ultrafast thermal shock of $Ti_3AlC_2$ ceramic in water and air, Int J Appl Ceram Technol 18 (2021) 1863–1871. https://doi.org/10.1111/ijac.13811.

[22] S. Li, H. Li, Y. Zhou, H. Zhai, Mechanism for abnormal thermal shock behavior of $Cr_2AlC$, Journal of the European Ceramic Society 34 (2014) 1083–1088. https://doi.org/10.1016/j.jeurceramsoc.2013.12.003.

[23] H. Li, S. Li, X. Du, P. Bao, Y. Zhou, Thermal shock behavior of $Cr_2AlC$ in different quenching media, Materials Letters 167 (2016) 131–133. https://doi.org/10.1016/j.matlet.2015.12.160.

[24] J.M. Schneider, Z. Sun, R. Mertens, F. Uestel, R. Ahuja, Ab initio calculations and experimental determination of the structure of $Cr_2AlC$, Solid State Communications 130 (2004) 445–449. https://doi.org/10.1016/j.ssc.2004.02.047.

[25] C. Walter, D.P. Sigumonrong, T. El-Raghy, J.M. Schneider, Towards large area deposition of $Cr_2AlC$ on steel, Thin Solid Films 515 (2006) 389–393. https://doi.org/10.1016/j.tsf.2005.12.219.

[26] J.D. Hettinger, S.E. Lofland, P. Finkel, T. Meehan, J. Palma, K. Harrell, S. Gupta, A. Ganguly, T. El-Raghy, M.W. Barsoum, Electrical transport, thermal transport, and elastic properties of $M_2AlC$ (M=Ti Cr, Nb, and V), Phys. Rev. B 72 (2005). https://doi.org/10.1103/PhysRevB.72.115120.

[27] B. Manoun, R.P. Gulve, S.K. Saxena, S. Gupta, M.W. Barsoum, C.S. Zha, Compression behavior of $M_2AlC$ (M=Ti V, Cr, Nb, and Ta) phases to above 50GPa, Phys. Rev. B 73 (2006). https://doi.org/10.1103/PhysRevB.73.024110.

[28] J. Wang, Y. Zhou, Dependence of elastic stiffness on electronic band structure of nanolaminate $M_2AlC$ (M=Ti,V,Nb and Cr ) ceramics, Phys. Rev. B 69 (2004). https://doi.org/10.1103/PhysRevB.69.214111.

[29] D. Music, Z. Sun, R. Ahuja, J.M. Schneider, Coupling in nanolaminated ternary carbides studied by theoretical means: The influence of electronic potential approximations, Phys. Rev. B 73 (2006). https://doi.org/10.1103/PhysRevB.73.134117.

[30] D. Music, A. Houben, R. Dronskowski, J.M. Schneider, Ab initio study of ductility in $M_2AlC$ (M=Ti V, Cr), Phys. Rev. B 75 (2007). https://doi.org/10.1103/PhysRevB.75.174102.

[31] P. Ström, D. Primetzhofer, Ion beam tools for nondestructive in-situ and in-operando composition analysis and modification of materials at the Tandem Laboratory in Uppsala, J. Inst. 17 (2022) P04011. https://doi.org/10.1088/1748-0221/17/04/P04011.

[32] M. Janson, CONTES Instruction Manual, Internal Report, Uppsala University, Sweden (2004).

[33] J.A. Leavitt, L.C. McIntyre, M.D. Ashbaugh, J.G. Oder, Z. Lin, B. Dezfouly-Arjomandy, Cross sections for 170.5° backscattering of $^4He$ from oxygen for $^4He$ energies between 1.8 and 5.0 MeV, Nuclear Instruments and Methods in Physics Research Section B: Beam Interactions with Materials and Atoms 44 (1990) 260–265. https://doi.org/10.1016/0168-583X(90)90637-A.

[34] M. Mayer, SIMNRA, a simulation program for the analysis of NRA, RBS and ERDA 541–544. https://doi.org/10.1063/1.59188.

[35] D. Nath, F. Singh, R. Das, X-ray diffraction analysis by Williamson-Hall, Halder-Wagner and size-strain plot methods of CdSe nanoparticles- a comparative study,

Materials Chemistry and Physics 239 (2020) 122021. https://doi.org/10.1016/j.matchemphys.2019.122021.

[36] W.C. Oliver, G.M. Pharr, An improved technique for determining hardness and elastic modulus using load and displacement sensing indentation experiments, J. Mater. Res. 7 (1992) 1564–1583. https://doi.org/10.1557/JMR.1992.1564.

[37] P. Hohenberg, W. Kohn, Inhomogeneous Electron Gas, Phys. Rev. 136 (1964) B864-B871. https://doi.org/10.1103/PhysRev.136.B864.

[38] W. Kohn, L.J. Sham, Self-Consistent Equations Including Exchange and Correlation Effects, Phys. Rev. 140 (1965) A1133-A1138. https://doi.org/10.1103/PhysRev.140.A1133.

[39] Kresse, Furthmüller, Efficient iterative schemes for ab initio total-energy calculations using a plane-wave basis set, Phys. Rev. B Condens. Matter 54 (1996) 11169–11186. https://doi.org/10.1103/PhysRevB.54.11169.

[40] G. Kresse, J. Furthmüller, Efficiency of ab-initio total energy calculations for metals and semiconductors using a plane-wave basis set, Computational Materials Science 6 (1996) 15–50. https://doi.org/10.1016/0927-0256(96)00008-0.

[41] Blöchl, Projector augmented-wave method, Phys. Rev. B Condens. Matter 50 (1994) 17953–17979. https://doi.org/10.1103/PhysRevB.50.17953.

[42] G. Kresse, D. Joubert, From ultrasoft pseudopotentials to the projector augmented-wave method, Phys. Rev. B Condens. Matter 59 (1999) 1758–1775. https://doi.org/10.1103/PhysRevB.59.1758.

[43] Perdew, Burke, Ernzerhof, Generalized Gradient Approximation Made Simple, Phys. Rev. Lett. 77 (1996) 3865–3868. https://doi.org/10.1103/PhysRevLett.77.3865.

[44] R. Yu, J. Zhu, H.Q. Ye, Calculations of single-crystal elastic constants made simple, Computer Physics Communications 181 (2010) 671–675. https://doi.org/10.1016/j.cpc.2009.11.017.

[45] R. Hill, The Elastic Behaviour of a Crystalline Aggregate, Proc. Phys. Soc. A 65 (1952) 349–354. https://doi.org/10.1088/0370-1298/65/5/307.

[46] R. Gaillac, P. Pullumbi, F.-X. Coudert, ELATE: an open-source online application for analysis and visualization of elastic tensors, J. Phys. Condens. Matter 28 (2016) 275201. https://doi.org/10.1088/0953-8984/28/27/275201.

[47] Y. Tian, B. Xu, Z. Zhao, Microscopic theory of hardness and design of novel superhard crystals, International Journal of Refractory Metals and Hard Materials 33 (2012) 93–106. https://doi.org/10.1016/j.ijrmhm.2012.02.021.

[48] X.-Q. Chen, H. Niu, D. Li, Y. Li, Modeling hardness of polycrystalline materials and bulk metallic glasses, Intermetallics 19 (2011) 1275–1281. https://doi.org/10.1016/j.intermet.2011.03.026.

[49] A. Togo, I. Tanaka, First principles phonon calculations in materials science, Scripta Materialia 108 (2015) 1–5. https://doi.org/10.1016/j.scriptamat.2015.07.021.

[50] A. Togo, L. Chaput, I. Tanaka, Distributions of phonon lifetimes in Brillouin zones, Phys. Rev. B 91 (2015). https://doi.org/10.1103/PhysRevB.91.094306.

[51] L. Chaput, Direct solution to the linearized phonon Boltzmann equation, Phys. Rev. Lett. 110 (2013) 265506. https://doi.org/10.1103/PhysRevLett.110.265506.

[52] G.K. Madsen, J. Carrete, M.J. Verstraete, BoltzTraP2, a program for interpolating band structures and calculating semi-classical transport coefficients, Computer Physics Communications 231 (2018) 140–145. https://doi.org/10.1016/j.cpc.2018.05.010.

[53] P.B. Allen, New method for solving Boltzmann's equation for electrons in metals, Phys. Rev. B Condens. Matter 17 (1978) 3725–3734. https://doi.org/10.1103/PhysRevB.17.3725.

[54] S. Poncé, E.R. Margine, C. Verdi, F. Giustino, EPW: Electron–phonon coupling, transport and superconducting properties using maximally localized Wannier functions, Computer Physics Communications 209 (2016) 116–133. https://doi.org/10.1016/j.cpc.2016.07.028.

[55] P. Giannozzi, S. Baroni, N. Bonini, M. Calandra, R. Car, C. Cavazzoni, D. Ceresoli, G.L. Chiarotti, M. Cococcioni, I. Dabo, A. Dal Corso, S. de Gironcoli, S. Fabris, G. Fratesi, R. Gebauer, U. Gerstmann, C. Gougoussis, A. Kokalj, M. Lazzeri, L. Martin-Samos, N. Marzari, F. Mauri, R. Mazzarello, S. Paolini, A. Pasquarello, L. Paulatto, C. Sbraccia, S. Scandolo, G. Sclauzero, A.P. Seitsonen, A. Smogunov, P. Umari, R.M. Wentzcovitch, QUANTUM ESPRESSO: a modular and open-source software project for quantum simulations of materials, J. Phys. Condens. Matter 21 (2009) 395502. https://doi.org/10.1088/0953-8984/21/39/395502.

[56] P. Giannozzi, O. Andreussi, T. Brumme, O. Bunau, M. Buongiorno Nardelli, M. Calandra, R. Car, C. Cavazzoni, D. Ceresoli, M. Cococcioni, N. Colonna, I. Carnimeo, A. Dal Corso, S. de Gironcoli, P. Delugas, R.A. DiStasio, A. Ferretti, A. Floris, G. Fratesi, G. Fugallo, R. Gebauer, U. Gerstmann, F. Giustino, T. Gorni, J. Jia, M. Kawamura, H.-Y. Ko, A. Kokalj, E. Küçükbenli, M. Lazzeri, M. Marsili, N. Marzari, F. Mauri, N.L. Nguyen, H.-V. Nguyen, A. Otero-de-la-Roza, L. Paulatto, S. Poncé, D. Rocca, R. Sabatini, B. Santra, M. Schlipf, A.P. Seitsonen, A. Smogunov, I. Timrov, T. Thonhauser, P. Umari, N. Vast, X. Wu, S. Baroni, Advanced capabilities for materials modelling with Quantum ESPRESSO, J. Phys. Condens. Matter 29 (2017) 465901. https://doi.org/10.1088/1361-648X/aa8f79.

[57] P. Giannozzi, O. Baseggio, P. Bonfà, D. Brunato, R. Car, I. Carnimeo, C. Cavazzoni, S. de Gironcoli, P. Delugas, F. Ferrari Ruffino, A. Ferretti, N. Marzari, I. Timrov, A. Urru, S. Baroni, Quantum ESPRESSO toward the exascale, J. Chem. Phys. 152 (2020) 154105. https://doi.org/10.1063/5.0005082.

[58] M. Dahlqvist, B. Alling, J. Rosén, Correlation between magnetic state and bulk modulus of $Cr_2AlC$, Journal of Applied Physics 113 (2013) 216103. https://doi.org/10.1063/1.4808239.

[59] Y.L. Du, Z.M. Sun, H. Hashimoto, M.W. Barsoum, Electron correlation effects in the MAX phase $Cr_2AlC$ from first-principles, Journal of Applied Physics 109 (2011) 63707. https://doi.org/10.1063/1.3562145.

[60] H.J. Monkhorst, J.D. Pack, Special points for Brillouin-zone integrations, Phys. Rev. B 13 (1976) 5188–5192. https://doi.org/10.1103/PhysRevB.13.5188.

[61] G. Prandini, A. Marrazzo, I.E. Castelli, N. Mounet, N. Marzari, Precision and efficiency in solid-state pseudopotential calculations, Computational Materials 4 (2018). https://doi.org/10.1038/s41524-018-0127-2.

[62] K. Lejaeghere, G. Bihlmayer, T. Björkman, P. Blaha, S. Blügel, V. Blum, D. Caliste, I.E. Castelli, S.J. Clark, A. Dal Corso, S. de Gironcoli, T. Deutsch, J.K. Dewhurst, I. Di Marco, C. Draxl, M. Dułak, O. Eriksson, J.A. Flores-Livas, K.F. Garrity, L. Genovese, P. Giannozzi, M. Giantomassi, S. Goedecker, X. Gonze, O. Grånäs, E.K.U. Gross, A. Gulans, F. Gygi, D.R. Hamann, P.J. Hasnip, N.A.W. Holzwarth, D. Iuşan, D.B. Jochym, F. Jollet, D. Jones, G. Kresse, K. Koepernik, E. Küçükbenli, Y.O. Kvashnin, I.L.M. Locht, S. Lubeck, M. Marsman, N. Marzari, U. Nitzsche, L. Nordström, T. Ozaki, L. Paulatto, C.J. Pickard, W. Poelmans, M.I.J. Probert, K. Refson, M. Richter, G.-M. Rignanese, S. Saha, M. Scheffler, M. Schlipf, K. Schwarz, S. Sharma, F. Tavazza, P. Thunström, A. Tkatchenko, M. Torrent, D. Vanderbilt, M.J. van Setten, V. van Speybroeck, J.M. Wills, J.R. Yates, G.-X.

Zhang, S. Cottenier, Reproducibility in density functional theory calculations of solids, Science 351 (2016) aad3000. https://doi.org/10.1126/science.aad3000.

[63] A.A. Mostofi, J.R. Yates, G. Pizzi, Y.-S. Lee, I. Souza, D. Vanderbilt, N. Marzari, An updated version of wannier90: A tool for obtaining maximally-localised Wannier functions, Computer Physics Communications 185 (2014) 2309–2310. https://doi.org/10.1016/j.cpc.2014.05.003.

[64] P. Ondračka, L. Löfler, NOMAD dataset: On the thermal shock behavior of MAX phase thin films - ab initio data, NOMAD Repository, 2022. https://doi.org/10.17172/NOMAD/2022.12.19-1.

[65] G. Greczynski, S. Mráz, L. Hultman, J.M. Schneider, Venting temperature determines surface chemistry of magnetron sputtered TiN films, Appl. Phys. Lett. 108 (2016) 41603. https://doi.org/10.1063/1.4940974.

[66] P. Finkel, M.W. Barsoum, T. El-Raghy, Low temperature dependencies of the elastic properties of $Ti_4AlN_3$, $Ti_3Al_{1.1}C_{1.8}$, and $Ti_3SiC_2$, Journal of Applied Physics 87 (2000) 1701–1703. https://doi.org/10.1063/1.372080.

[67] G. Ying, X. He, M. Li, W. Han, F. He, S. Du, Synthesis and mechanical properties of high-purity $Cr_2AlC$ ceramic, Materials Science and Engineering: A 528 (2011) 2635–2640. https://doi.org/10.1016/j.msea.2010.12.039.

[68] H. Li, H. Cao, F. Liu, Y. Li, F. Qi, X. Ouyang, N. Zhao, Microstructure, mechanical and electrochemical properties of $Ti_3AlC_2$ coatings prepared by filtered cathode vacuum arc technology, Journal of the European Ceramic Society 42 (2022) 2073–2083. https://doi.org/10.1016/j.jeurceramsoc.2021.12.066.

[69] J.M. Schneider, D.P. Sigumonrong, D. Music, C. Walter, J. Emmerlich, R. Iskandar, J. Mayer, Elastic properties of $Cr_2AlC$ thin films probed by nanoindentation and ab initio molecular dynamics, Scripta Materialia 57 (2007) 1137–1140. https://doi.org/10.1016/j.scriptamat.2007.08.006.

[70] B. Völker, B. Stelzer, S. Mráz, H. Rueß, R. Sahu, C. Kirchlechner, G. Dehm, J.M. Schneider, On the fracture behavior of $Cr_2AlC$ coatings, Materials & Design 206 (2021) 109757. https://doi.org/10.1016/j.matdes.2021.109757.

[71] J.S.K.-L. Gibson, S. Rezaei, H. Rueß, M. Hans, D. Music, S. Wulfinghoff, J.M. Schneider, S. Reese, S. Korte-Kerzel, From quantum to continuum mechanics: studying the fracture toughness of transition metal nitrides and oxynitrides, Materials Research Letters 6 (2018) 142–151. https://doi.org/10.1080/21663831.2017.1414081.

[72] P. Zhang, S.X. Li, Z.F. Zhang, General relationship between strength and hardness, Materials Science and Engineering: A 529 (2011) 62–73. https://doi.org/10.1016/j.msea.2011.08.061.

[73] W. Skrotzki, A. Pukenas, E. Odor, B. Joni, T. Ungar, B. Völker, A. Hohenwarter, R. Pippan, E.P. George, Microstructure, Texture, and Strength Development during High-Pressure Torsion of CrMnFeCoNi High-Entropy Alloy, Crystals 10 (2020) 336. https://doi.org/10.3390/cryst10040336.

[74] W.D. Callister, Materials science and engineering: an introduction, John Wiley & Sons, 2006.ISBN: 978-0-471-73696-7.

[75] J.F. Shackelford (Ed.), CRC materials science and engineering handbook, 3rd ed.CRC Press, 2001. ISBN: 0-8493-2696-6

[76] N.V. Tzenov, M.W. Barsoum, Synthesis and Characterization of $Ti_3AlC_2$, Journal of the American Ceramic Society 83 (2000) 825–832. https://doi.org/10.1111/j.1151-2916.2000.tb01281.x.

[77] G. Ying, X. He, M. Li, S. Du, W. Han, F. He, Effect of $Cr_7C_3$ on the mechanical, thermal, and electrical properties of $Cr_2AlC$, Journal of Alloys and Compounds 509 (2011) 8022–8027. https://doi.org/10.1016/j.jallcom.2011.04.134.

[78] D.T. Wan, F.L. Meng, Y.C. Zhou, Y.W. Bao, J.X. Chen, Effect of grain size, notch width, and testing temperature on the fracture toughness of $Ti_3Si(Al)C_2$ and Ti3AlC2 using the chevron-notched beam (CNB) method, Journal of the European Ceramic Society 28 (2008) 663–669. https://doi.org/10.1016/j.jeurceramsoc.2007.07.011.

[79] S. Karimi Aghda, D. Music, Y. Unutulmazsoy, H.H. Sua, S. Mráz, M. Hans, D. Primetzhofer, A. Anders, J.M. Schneider, Unravelling the ion-energy-dependent structure evolution and its implications for the elastic properties of (V,Al)N thin films, Acta Materialia 214 (2021) 117003. https://doi.org/10.1016/j.actamat.2021.117003.

[80] D. Eichner, A. Schlieter, C. Leyens, L. Shang, S. Shayestehaminzadeh, J.M. Schneider, Solid particle erosion behavior of nanolaminated $Cr_2AlC$ films, Wear 402-403 (2018) 187–195. https://doi.org/10.1016/j.wear.2018.02.014.

[81] N. Li, Y. Mo, W.-Y. Ching, The bonding, charge distribution, spin ordering, optical, and elastic properties of four MAX phases $Cr_2AX$ (A = Al or Ge, X = C or N): From density functional theory study, Journal of Applied Physics 114 (2013) 183503. https://doi.org/10.1063/1.4829485.

[82] Z. Sun, S. Li, R. Ahuja, J.M. Schneider, Calculated elastic properties of $M_2AlC$ (M=Ti, V, Cr, Nb and Ta), Solid State Communications 129 (2004) 589–592. https://doi.org/10.1016/j.ssc.2003.12.008.

[83] W. Tian, P. Wang, G. Zhang, Y. Kan, Y. Li, D. Yan, Synthesis and thermal and electrical properties of bulk $Cr_2AlC$, Scripta Materialia 54 (2006) 841–846. https://doi.org/10.1016/j.scriptamat.2005.11.009.

[84] T. Scabarozi, A. Ganguly, J.D. Hettinger, S.E. Lofland, S. Amini, P. Finkel, T. El-Raghy, M.W. Barsoum, Electronic and thermal properties of $Ti_3Al(C_{0.5},N_{0.5})_2$,

$Ti_2Al(C_{0.5},N_{0.5})$ and $Ti_2AlN$, Journal of Applied Physics 104 (2008) 73713. https://doi.org/10.1063/1.2979326.

[85] A. Champagne, J.-L. Battaglia, T. Ouisse, F. Ricci, A. Kusiak, C. Pradere, V. Natu, A. Dewandre, M.J. Verstraete, M.W. Barsoum, J.-C. Charlier, Heat Capacity and Anisotropic Thermal Conductivity in $Cr_2AlC$ Single Crystals at High Temperature, J. Phys. Chem. C 124 (2020) 24017–24028. https://doi.org/10.1021/acs.jpcc.0c08384.

[86] X. Wang, Y. Zhou, Microstructure and properties of $Ti_3AlC_2$ prepared by the solid–liquid reaction synthesis and simultaneous in-situ hot pressing process, Acta Materialia 50 (2002) 3143–3151. https://doi.org/10.1016/S1359-6454(02)00117-9.

[87] N.J. Lane, S.C. Vogel, E.N. Caspi, M.W. Barsoum, High-temperature neutron diffraction and first-principles study of temperature-dependent crystal structures and atomic vibrations in $Ti_3AlC_2$ $Ti_2AlC$, and Ti 5 Al 2 C 3, Journal of Applied Physics 113 (2013) 183519. https://doi.org/10.1063/1.4803700.

[88] L. Peng, Fabrication and properties of $Ti_3AlC_2$ particulates reinforced copper composites, Scripta Materialia 56 (2007) 729–732. https://doi.org/10.1016/j.scriptamat.2007.01.027.

[89] Y.W. Bao, C.F. Hu, Y.C. Zhou, Damage tolerance of nanolayer grained ceramics and quantitative estimation, Materials Science and Technology 22 (2006) 227–230. https://doi.org/10.1179/174328406X86209.

[90] S.B. Li, W.B. Yu, H.X. Zhai, G.M. Song, W.G. Sloof, S. van der Zwaag, Mechanical properties of low temperature synthesized dense and fine-grained $Cr_2AlC$ ceramics, Journal of the European Ceramic Society 31 (2011) 217–224. https://doi.org/10.1016/j.jeurceramsoc.2010.08.014.

[91] W. Tian, P. Wang, G. Zhang, Y. Kan, Y. Li, Mechanical Properties of $Cr_2AlC$ Ceramics, Journal of the American Ceramic Society 90 (2007) 1663–1666. https://doi.org/10.1111/j.1551-2916.2007.01634.x.

[92] W.B. Zhou, B.C. Mei, J.Q. Zhu, On the synthesis and properties of bulk ternary $Cr_2AlC$ ceramics, Materials Science-Poland 27 (2009) 973–980.

[93] T.H. Scabarozi, S. Amini, O. Leaffer, A. Ganguly, S. Gupta, W. Tambussi, S. Clipper, J.E. Spanier, M.W. Barsoum, J.D. Hettinger, S.E. Lofland, Thermal expansion of select $M_{n+1}AX_n$ (M=earlytransitionmetal, A=Agroupelement, X=C or N) phases measured by high temperature x-ray diffraction and dilatometry, Journal of Applied Physics 105 (2009) 13543. https://doi.org/10.1063/1.3021465.

[94] T. Cabioch, P. Eklund, V. Mauchamp, M. Jaouen, M.W. Barsoum, Tailoring of the thermal expansion of $Cr_2(Al_x,Ge_{1-x})C$ phases, Journal of the European Ceramic Society 33 (2013) 897–904. https://doi.org/10.1016/j.jeurceramsoc.2012.10.008.